\documentstyle[aps,preprint,floats,epsf]{revtex}
\def\be{\begin{equation}}
\def\ee{\end{equation}}
\def\ba{\begin{eqnarray}}
\def\ea{\end{eqnarray}}
\def\bml{\begin{mathletters}}
\def\eml{\end{mathletters}}
\def\nn{\nonumber\\}
\begin{document}
\draft

%
%
%
\renewcommand{\topfraction}{0.8}
\renewcommand{\bottomfraction}{0.8}

\title{$G_1$ Cosmologies with Gravitational and Scalar Waves} 
\author{Ruth Lazkoz}    
\address{Astronomy Unit, School of Mathematical Sciences
\\Queen Mary \& Westfield College, London E1 4NS U.K. \\
and
\\Fisika Teorikoaren Saila\\
Euskal Herriko Unibertsitatea\\
644 P.K., 48080 Bilbao, Spain
}  
\date{\today} 
\maketitle
\begin{abstract}
I present here a new algorithm to generate families of inhomogeneous 
massless scalar field cosmologies.
New spacetimes, having a single isometry, are generated
 by breaking the homogeneity of 
massless scalar field $G_2$ models along one direction.
As an illustration of the technique I construct cosmological models
which in their late time limit represent  perturbations in the form of 
gravitational and scalar waves propagating on a non-static
inhomogeneous background. Several features of the obtained 
metrics are discussed, such as their early and late time limits, 
structure of singularities and physical interpretation.
\end{abstract}



\section{Introduction}  
The high degree of isotropy observed in the Universe on large
scales today is usually combined with the Copernican principle 
to justify the assumption of homogeneity on the same scales. 
However, there is no known reason to assume that the Universe 
was isotropic nor homogeneous at very early epochs.
The puzzling question of how the Universe might have evolved 
from an initially irregular state into the current isotropic and 
apparently homogeneous state lacks a complete answer 
at present. To date, several regularization mechanisms have 
been put forward, such as Misner's chaotic cosmological program 
\cite{misner1,misner2}, the standard inflationary scenario
 \cite{infla1,infla2,infla3,infla4} and, more recently, the
alternative pre-Big Bang inflationary scenario \cite{pbb}.
However, none of these is completely satisfactory, and 
in general one cannot know for certain which range of initial 
conditions could have allowed
 the Universe evolve into its present form.
Such scenarios can only provide one with partial indications 
of what initial
conditions would have led a generic universe into the one observed
at present.  One way to maximize the amount of information
obtainable from any cosmological program, such as those mentioned
above, relies on studying the evolution of models with as many degrees
of freedom as possible. This idea has motivated the special attention paid to 
inhomogeneous cosmological models 
in the last decades (see Krasinski \cite{kra} for a review).

In general,  attempts  to obtain new inhomogeneous metrics 
involve some symmetry assumption, so that
the full implications of the non-linearity of
the theory are compromised to some extent.
In addition, non-vacuum spacetimes  are required 
to have a physically meaningful matter content,
 so that they portray realistic situations. 
Motivated by the possibility of the
existence of non-trivial massless scalar fields in the early universe,
I will concern myself here with cosmological solutions to
the Einstein equations induced by such matter sources. In particular,
I present a new algorithm to generate 
families of massless scalar field $G_1$ cosmologies, i.e. 
time-dependent spacetimes with a  single isometry.
These new sets of metrics will be generated starting from
generalized vacuum Einstein-Rosen spacetimes, 
which admit an Abelian group of isometries
$G_2$ acting transitively on spacelike surfaces.
In the last decades there has been 
intensive study of  $G_2$ vacuum and matter filled cosmological
models  and several major reviews on the subject 
have been written \cite{mac,kramer,kra}.
  
Given the large number of known $G_2$ cosmologies and the various
available techniques to generate further ones, 
algorithms transforming such
spacetimes into $G_1$ metrics represent powerful
tools for generating new inhomogeneous solutions.
The new algorithm, which I shall present, displays the nice
property of reducing the symmetry while 
keeping the type of matter source unaltered. 
Nonetheless, the input and output solutions
may not admit the same physical interpretation
or even share some of the same relevant features. 
For this reason, if one wishes to grasp the physical
meaning of every new solution, an independent analysis of it
will have to be carried out.

Here in I construct and analyze $G_1$ models representing
the propagation of gravitational and matter waves
on a non-spatially flat background.  
The study of primordial inhomogeneities in the form of waves 
is an active area of research. This is  
motivated by the fact that  wave-like primordial perturbations
originating from vacuum fluctuations 
during inflation may be responsible for structure formation.
Unlike other types of inhomogeneities formed in the early universe, 
they would have remained nearly unaltered up to the present, 
and therefore allowing the possibility of their detection.

The exact inhomogeneous $G_1$ spacetimes seeded by gravitational 
and matter waves studied here represent a generalization of 
more symmetric configurations considered by Charach and Malin \cite{cha};
in those models the background hosting the waves was homogeneous. 
In the context of colliding plane waves,
$G_2$ diagonal spacetimes with waves of scalar and gravitational 
nature have also been considered. Space-times such as those 
studied by Wu \cite{wu} or any solution generated by the methods of
Barrow \cite{bar} and Wainwright et al.  \cite{wim} could  be taken 
as starting points to construct  new $G_1$ metrics modeling 
interactions of waves on a curved background. 

Furthermore, interest in this generation procedure is not  
restricted to the relativistic framework; solutions 
to Einstein's equations with a massless scalar field (in what follows
m.s.f.) may be used
to generate solutions to alternative theories of gravity, such as
Brans-Dicke theory or string theory in its low energy limit.
In the latter case, one could even take those spacetimes
to generate new solutions with other massless modes 
in the characteristic spectrum of the theory.

The plot of the paper is as follows: First I introduce the $G_1$
massless scalar field solution generating algorithm itself. 
Then, I construct new inhomogeneous
metrics starting from a infinite dimensional family of solutions which 
in the WKB limit admit an interpretation in terms of waves.
It will be shown that at early times these solutions behave
like a Belinskii-Khalatnikov generalized model \cite{bk}
with homogeneity broken along one 
spatial direction. The structure of spacelike 
singularities of the new solutions in the early time limit 
will be analyzed as well, and the special features due to 
the presence of the matter source will be indicated.
Next I will consider the solutions' high frequency limit and show 
that they can be thought of in the same physical terms as their $G_2$
counterparts.
In particular the new cosmological models represent a spacetime with a
spatially inhomogeneous background filled with a non-static
inhomogeneous  scalar field and a null fluid of ``gravitons" and
 ``scalar particles." As time 
grows the null fluid's contribution to the energy-momentum tensor 
grows faster than that associated with the homogeneous 
part of the scalar field generating the  background geometry.
Thus, the model evolves into an inhomogeneous generalization of
the cosmological model of Doroshkevich, Zeldovich and Novikov (DZN)
 \cite{dzn}. Finally, the main conclusions are outlined.

\section{Solution Generation Algorithm}
Basically, the new generating technique is a prescription
to break homogeneity along one direction in $G_2$ 
m.s.f. cosmologies, ending up with new spacetimes possessing a
single isometry but having the same type of matter content. 
Remarkably, the pioneering investigations on matter filled universes of 
Einstein-Rosen type, carried out respectively  by Tabensky and Taub
\cite{tab} and Liang \cite{liang} considered models 
with non interacting scalar fields, even though 
at the time there was no clear physical motivation.
 
According to a conjecture due to Belinskii, Lifshitz and Khalatnikov (BLK)
 \cite{blk1,blk2,blk3,blk4,blk5,blk6},
$G_2$ metrics seem to be specially relevant 
for the description of the early universe,
as such solutions give the leading approximation 
to a general solution near the singularity at $t=0$.
Their claim has very recently found the support of numerical results 
\cite{ber}. In particular BLK considered approximate 
Einstein-Rosen solutions and  performed an analysis of their 
local behavior in the early and late time regime. An interesting result 
reached in the course of those investigations, which is specially relevant
for the present paper,  was the prediction of a high frequency
gravitational wave regime in the late epochs of the Universe.

Before going any further it is convenient to explain
how  $G_2$ spacetimes  induced by a m.s.f.
can be generated starting from vacuum solutions to 
Einstein's equations with the same symmetry. 
For the sake of simplicity, the discussion will be 
restricted here to a particular case of a well-known general procedure
\cite{lete,cha,charev,bar,wim,ccf,verda}. 
The generic diagonal line-element with $G_2$
symmetry will be taken as a starting point:
\be
 ds_v^2=e^{f_v(t,z)}(-dt^2+dz^2)+
G_v(t,z)\left(e^{p_v(t,z)}dx^2+e^{-p_v(t,z)}dy^2\right)\,,
\ee
where the subscript $v$ stands for vacuum.
A new solution $g_{\mu\nu}$ of the Einstein equations with a massless
scalar field $\varphi$ as a source and line-element
\be
 ds^2=e^{f(t,z)}(-dt^2+dz^2)+
G(t,z)\left(e^{p(t,z)}dx^2+e^{-p(t,z)}dy^2 \right) \label{g2}\,,
\ee
can be obtained by the following transformations:
\bml
\label{g2bis}
\ba
G&=&G_v\,,\\
p&=&B \,p_v+C\, \log G_v\,, \\
f&=&f_v+E \,p_v+F\, \log G_v\,,\\
\varphi&=&A\,p_v+D\,\log G_v\,;\ea
\eml
provided that the constants $A,\,B,\,C,\,D,\,E$ and $F$ 
are subject to the constraints:
\bml
\label{condit}
\ba
&\,&B\,C+2\,A\,D=E\,, \\
 &\,&C^2+2\,D^2=2\,F\,,\\
&\,&B^2+2\,A^2=1\,.
\ea
\eml
The conditions (\ref{condit}) arise by demanding 
the following are satisfied:
\bml
\ba&R_{\mu\nu}=\varphi_{,\mu}\varphi_{,\nu}&\,,\\
&g^{\mu\nu}\nabla_{\mu}\nabla_{\nu}
 \varphi=0&\,.\ea\eml

In principle a large number of new 
m.s.f. $G_2$ cosmologies can be obtained by simply applying 
the procedure sketched above to any of the representatives of 
the populated family of vacuum Einstein-Rosen spacetimes.
However, generating m.s.f. solutions with a lower degree of 
symmetry is a more cumbersome task.  At this point I would like to 
draw attention to a method given by Feinstein et al. \cite{vos}
which allows one to generate families of solutions with a
two-dimensional degree of inhomogeneity and 
a self-interaction term for the massless field of  the form
\be V=V_0(\lambda)\,e^{-\lambda \,\varphi}.\ee 
With these procedure new metrics are obtained by introducing an 
$x\hbox{-dependent}$ conformal factor on the input $G_2$ metric, 
and where in general the potential term only vanishes
for $\vert \lambda \vert=6$. This difficulty 
in canceling the self-interaction term for the scalar field 
can be traced back to the high degree of symmetry of the 
$x\hbox{-dependence}$ in the models considered in ref. \cite {vos}. 

In order to find a prescription for introducing an 
additional degree of symmetry in m.s.f. $G_2$ metrics without 
switching on a potential in the process, I  have considered 
the possibility of having a more general $x\hbox{-dependence}$ in the 
metric. In particular, I have sought metrics of the form:
\be
d\tilde s^2=\Omega(x)\,e^{f(t,z)}(-dt^2+dz^2)+
G(t,z)\left(e^{p(t,z)}dx^2+\Xi (x)\,e^{-p(t,z)}dy^2\right)\label{gen}\,,
\ee
and made the following ansatz for the  scalar field:
\be \tilde \varphi (t,z,x)= \varphi (t,z)+\Lambda (x)\,.\ee
Note that the massless scalar field case included in the solutions
of Feinstein et al. is also a particular case of the models here. 
Since the requirement is that no potential should arise
in the transformations, the equations that must hold are:
\bml\ba
\label{ric}
 &\tilde R_{\mu\nu}=\tilde\varphi_{,\mu}
\tilde\varphi_{,\nu}&\label{ricci}\,,\\
 &\tilde g^{\mu\nu}\tilde \nabla_{\mu}
\tilde \nabla_{\nu}\tilde\varphi=0\,.& \label{klein} \ea\eml

Explicitly, equation (\ref{ricci}) is equivalent to the set of equations:
\bml 
\label{r}
\ba
\tilde R_{00}&=&R_{00}+
 {e^{f - p}}\, {\frac{
      \Xi_{,x}\,\Omega _{,x} + 2\,\Xi \,\Omega _{,xx}}{4\,
      G\,\Xi}} =\tilde\varphi_{,t}^2 \label{r0}\,,\\
\tilde R_{11}&=& R_{11}-{{e^{f- p}}\,\frac{ 
        \Xi _{,x}\,\Omega _{,x} + 2\,\Xi\,\Omega _{,xx}
         }{4\,G\,\Xi}}=\tilde\varphi_{,z}^2\,,\\
\tilde R_{01}&=&R_{01}=\tilde\varphi_{,t}\tilde\varphi_{,z}
\,,\\
\tilde R_{22}&=&R_{22}+
{\frac{{{\Xi_{,x}}^2} - 2\,\Xi\,\Xi _{,xx}}{4\,{{\Xi}^2}}}+
{\frac{{{\Omega_{,x}}^2}
 - 2\,\Omega\,\Omega _{,xx}}
    {2\,{{\Omega}^2}}} =\tilde\varphi_{,x}^2
\,,\\
\tilde R_{12}&=&\frac{G_{,z}\,\Omega_{,x}}{2\,G\,\Omega} + 
{\frac{\Xi _{,x}\,p_{,z}}{2\,\Xi}}=\tilde\varphi_{,z}\tilde\varphi_{,x}\,,\\
\tilde R_{02}&=&{{{\frac{G_{,t}\,\Omega_{,x}}{2\,G\,\Omega}} + 
     {\frac{\Xi_{,x}\,p_{,t}}{2\,\Xi}}}}=
\tilde\varphi_{,t}\tilde\varphi_{,x}\,,\\
\tilde R_{33}&=&R_{33}+
{e^{-2\,p}}{\frac{
      \, {{\Xi _{,x}}^2} -2\,\Xi\,\Xi _{,xx}}{4\,
      \,\Xi }}-
{e^{-2\,p}}{\frac{ \,\Xi _{,x}\,\Omega _{,x}}{2\,\Omega}}=0\label{r3}\,;
\ea
\eml
whereas  equation (\ref{klein})  in explicit form reads
\be \frac{g^{\mu\nu}\nabla_{\mu}\nabla_{\nu} \varphi}{\Omega}+
e^{-p}\frac{(\Omega\,\Xi^{1/2}\tilde\varphi_{,x})_{,x}}{\Xi^{1/2}}=0\,. 
\ee
Inspection of the equations (\ref{r}) indicates that the three
$x\hbox{-dependent}$ metric functions must have the form:
\bml
\label{ansatz}
\ba\Omega(x)&=&x^k\,,\\
\Xi(x)&=&x^n\,,\\
\Lambda(x)&=&m\log \vert x \vert\,;\ea\eml 
subject to the following constraints on the parameters $k$, $n$ and $m$:
\bml
\label{esas}
\ba &\,&k\,(2\,k+n-2)=0\,,\label{e1}\\
&\,&n\,(2\,k+n-2)=0\,,\label{e2}\\
&\,&k+n\,C=2\,D\,m\,,\\
&\,&n\,B=2\,A\,m\,,\\
&\,&2\,m^2=4\,k-3\,k^2\,.\label{e3}\ea\eml
In this case, the Klein-Gordon equation (\ref{klein}) reduces to 
\be \label{kg} m\,\,(2\,k+n-2)=0\,, \label{esta}\ee which is automatically
satisfied provided that (\ref{e1},\ref{e2}) and (\ref{e3}) hold. 
Note that the parameter $k$ must  be non-negative and not larger than
$4/3$. The $x\hbox{-dependent}$ term
of the scalar field will be maximum for $k=2/3$. Consistency of the 
solutions is reflected by the fact that for the following choice: 
\be
m=n=k=0\,.
\ee
the set of equations (\ref{esas},\ref{esta}) is satisfied for any value of 
$A$, $B$, $C$ and $D$, and the input $G_2$ m.s.f is recovered. 
On the one  hand, for  $m\ne0$ and $k\ne 1$, if $k$ and $C$ are 
taken as free parameters,
one can parameterize the solution's constants in the form:
\bml
\ba n&=&2-2\,k\,,\\
m&=&\hbox{sign}\,(m)\frac{\sqrt{4\,k-3\,k^2}}{\sqrt{2}}\,,\\
A&=&\sqrt{2}\,\hbox{sign}\,(A)\left\vert\frac{1-k}{2-k }\right\vert\,,\\
B& =&\hbox{sign}{\,(A\,m)}\left\vert\frac{1-k}{
2-k}\right\vert
\frac{\sqrt{4\,k-3\,k^2}}{1-k}\,,\\
D&=&\hbox{sign}\,(m)\frac{k+(2-2\,k)\,C}{\sqrt{8\,k-6\,k^2}}\,,\\
E&=&\hbox{sign}{\,(A\,m)}\left\vert\frac{2-2\,k}{2-k}\right\vert
\frac{C\,(2-k)^2+(2-2\,k)\,k}{\sqrt{4\,k-3\,k^ 2}}\,,
\\
F&=&\frac{C^2}{2}+\frac{\left(k+(2-2\,k)\,C\right)^2}{8\,k-6\,k^2}\,.
\ea\eml
Four different subcases can be distinguished, depending on the
choice of the sign of $A$ and $m$.
On the other hand, in the particular case 
$m\ne0$, $k=1$, the constants take the values:
\bml
\ba
&&\sqrt{2}\vert m \vert =\sqrt{2}\vert D\vert=\vert B\vert\\
&&A=0\\
&&C^2+1=2\,F\\
&&E=\hbox{sign}{\,(B)}
\ea
\eml

In general the metric $\tilde g_{\mu\nu}$ admits only one Killing and 
one homothetic vector,  namely:
\bml\ba
\xi_{kv}&=&\frac{\partial }{\partial y}\\
\xi_{hv}&=&y\frac{\partial}{\partial y},
\ea\eml
where by definition: 
\bml\ba
{\cal L}_{\xi_{ kv}}g_{\mu\nu}&=&0,\\
{\cal L}_{\xi_{hv}}g_{\mu\nu}&=&2g_{\mu\nu};
\ea\eml

It is known that a matter source and the geometry it induces need not
share the same symmetry properties unless it is a perfect fluid 
\cite{coley,carr}. This property  is usually referred to as the
 ``inheritance problem''. Inspection of the metrics obtained here
show that for $k=4/3$ the $x\hbox{-dependent}$ term of the scalar field 
is absent, even though the metric depends on that coordinate in a 
non-trivial way. This is equivalent to:
\bml\ba
{\cal L}_{\xi_{mc}}T_{\mu\nu}&=&0\\
{\cal L}_{\xi_{mc}}g_{\mu\nu}&\ne&0
;
\ea
\eml
where 
$${\xi_{mc}}=\frac{\partial}{\partial x}$$
and the vector ${\xi_{mc}}$ 
is a so-called matter collineation \cite{kramer,carot,carr}.

Bearing in mind the similarity between the new algorithm and the one
of \cite{vos}, one might wonder whether geometries like (\ref{gen})
 can be also seeded by a scalar field with an exponential potential. 
Such a situation may be shown to be only possible if $n=k$, which 
is nothing but the case already found by Feinstein et al. 
In order to prove this, let us consider the  case where
the generic geometry (\ref{gen}), under the constraint (\ref{ansatz}), 
is induced by an  exponential potential  
$\tilde V(\tilde \varphi)=V_0(\lambda)\, 
e^{-\lambda \tilde \varphi}$. In this case the field equations are:
\bml
\ba
&&\tilde R_{\mu\nu}=\tilde\varphi_{,\mu}\tilde\varphi_{,\nu}+\tilde
g_{\mu\nu}V_0\, e^{-\lambda
\tilde \varphi}\,,\\
&&\tilde\nabla^{\mu}\tilde\nabla_{\mu}\tilde\varphi=-
\frac{\partial \tilde V (\tilde \varphi)}{\tilde \partial 
\tilde \varphi}\,.
\ea
\eml
It is only necessary to look at the equations for $\tilde R_{00}$ and
$\tilde R_{33}$ to realize the following constraint must hold:
 \be 
{k\,(2-2\,k-n)}={n\,(2-2\,k-n)}=4\,V_0\,\ne 0.
\ee 
Compatibility of the latter set of equations in the case of a non vanishing
potential requires $n=k$, or in other words, that the $x\hbox{-dependence}$ 
of the metric is given by a global conformal factor 
as in the case studied by
\cite{vos}.

It is important to note here that the generation technique does not
restrict the character of the gradient of metric function $G(t,z)$ 
of the input metric. That function determines the local
behavior of the spacetime, and its gradient can be globally timelike,
spacelike, null or vary from point to point. Although in what
follows I am focusing on a case 
with a timelike character of the gradient of $G(t,z)$, a window
is left open for the study of other physically appealing cases.

Moreover, even though our generating prescription
has been used to break homogeneity of an input m.s.f. solution
with  a single degree of inhomogeneity, 
it is also possible to construct an equivalent algorithm
that would transform certain static spacetimes into  
non-static ones. One should start with a m.s.f. model
with two commuting Killing vectors, one of them being timelike,
and then generalize it by introducing time dependent factors
in the metric and scalar field as I have done here.

\section{Cosmologies with gravitational and scalar waves}

After having outlined our method to generate uniparametric
families of $G_1$ cosmologies, I shall now construct
the counterparts of a family of $G_2$ cosmologies 
found by Charach and Malin \cite{cha}. That set of metrics,
which can be thought of as inhomogeneous sinusoidal 
perturbations of the well-known  Bianchi I spacetimes,
represent the propagation of gravitational and scalar waves on an 
anisotropically expanding flat background. Simpler models 
of that sort, not including scalar degrees of freedom, 
were studied earlier on by Berger \cite{ber}.

In this paper I  am  only considering Charach and Malin's solutions in
the asymptotic limits $t\sim 0$ and $j\,t\gg1$ for any value of $j$.
It is convenient, however, to 
give here the full expressions of the metric functions of the vacuum
solution from which they were derived, namely:
\bml
\label{full}
\ba
G_v&=&t\,,\\
p_v&=&\beta \log t +\sum_{j=1}^{\infty} \cos [j(z-z_n)]\alpha_j J_0(j\,t)\,,\\
f_v&=&\frac{\beta^2-1}{2}+\beta\sum_{j=1}^{\infty}\alpha_j\cos[\alpha_j(z-z_j)]
J_0(j\,t)+\frac{t^2}{4}\sum_{j=1}^{\infty}j^2\{[\alpha_jJ_0(j\,t)]^2+
[\alpha_jJ_1(j\,t)]^2\}\nn
&-&\frac{t}{2}\sum_{j=1}^{\infty}j\alpha_j^2\cos^2[\alpha_j(z-z_j)]
J_0(j\,t)J_1(j\,t)
+\frac{t}{2}\sum_{l=1}^{\infty}\sum_{j=1}^{\infty}\frac{l\,j}{l^2-j^2}\nn
&\times&\{\sin[l(z-z_l)]\sin[j(z-z_j)]
\left[l\alpha_l\alpha_jJ_1(l\,t)J_0(j\,t)-j\alpha_l\alpha_jJ_0(l\,t)J_1(j\,t)\right]\nn
&+&\cos[l(z-z_l)]\cos[j(z-z_j)]
\left[j\alpha_l\alpha_jJ_1(l\,t)J_0(j\,t)\right.
-\left.l\alpha_l\alpha_jJ_0(l\,t)J_1(j\,t)\right]\}\,.\ea
\eml

Charach and Malin's solutions were obtained 
by breaking the spatial homogeneity of 
Belinskii-Khalatnikov  homogeneous solution 
\cite{bk}, which also has a m.s.f. as a seed. 
The gravitational and scalar degrees of freedom  of those spacetimes
 satisfy linear wave equations, with 
the form of cylindrically symmetric waves propagating on Minkowski 
spacetime, for which the spatial and temporal coordinates 
have been interchanged.
Since the solutions are of the standing wave form they are 
fully compatible with the $S^1\otimes S^1\otimes S^1$ topology 
(three-torus). However, the  $G_1$ counterparts of those 
spacetimes cannot have the same topology,
$x$ cannot be a cyclic coordinate in this case due to the presence
of the term proportional to $\log|x|$ in the scalar field $\tilde \varphi$.

\subsection{Early time behavior and singularities}
\label{early}
From the analysis of  the new $G_1$ solution's  early time 
behavior it can be  determined whether spacelike singularities arise 
at the time origin $t=0$. 
A look at expressions (\ref{full}) shows that in the case under
discussion, the periodic inhomogeneities can be neglected
in the very first stages of that spacetime's evolution.
Our metrics will be then $G_2$ inhomogeneous generalizations 
of the cosmological models of Doroshkevich-Zeldovich-Novikov
(DZN) \cite{dzn}, which have three commuting Killing vectors.
In this limit  the metric and scalar field read
\bml
\ba
&&\tilde g_{\mu\nu}\sim\hbox{diag}
\left(-x^k\epsilon_1(t),x^k\epsilon_1(t),
\epsilon_2(t),x^n\epsilon_3(t)\right)\label{bel}\\
&&\tilde\varphi=\varphi_0\log t+m\log \vert x \vert
\ea
\eml
where
\bml
\ba
&&\epsilon_1=t^{f_0}\,,\\
&&\epsilon_2=t^{1+p_0}\,,\\
&&\epsilon_3=t^{1-p_0}\,,
\ea
\eml
and
\bml
\label{cons}\ba 
p_0&=&B\beta+C\,,\\
f_0&=&\frac{\beta^2-1}{2}+E\beta+F\,,\\
\varphi_0&=&A\beta+D\,.
\ea\eml
In addition, the following relationship holds:
\be
f_0=\frac{p_0^2-1}{2}+\varphi_0^2\,.\label{condi}
\ee
Following Charach and Malin, the metric 
can be rewritten using a synchronous set of 
coordinates in which the new time coordinate $\tau$ is defined
\be d\tau=\sqrt{\epsilon_1(t)}dt\,;\ee
this way the metric can be recognized as a simple inhomogeneous
generalization of a Belinskii-Khalatnikov \cite{bk} solution.

In broad terms, the presence of a spacelike singularity
at early times will be reflected in the behavior 
of the curvature invariants $R$, $R_{\mu\nu}R^{\mu\nu}$ and
 $R_{\mu\nu\sigma\delta}R^{\mu\nu\sigma\delta}$.
Due to the inhomogeneous character of the metric 
it is possible in principle to have a conspiracy between the
parameters, so that on certain hypersurfaces some those invariants
are identically null and therefore do not reveal  
the presence of a singularity in the spacetime.
Since the Kretchmann scalar 
$K=R_{\mu\nu\sigma\delta}R^{\mu\nu\sigma\delta}$ 
is however non-null and positive everywhere 
I  will use this in the search for singularities. In the  case 
here it explicitly reads:
\ba
K&=&\frac{3\,f_0^2}{8\,t^{6\,f_0+4}x^{6\,k}}
\left[(1+p_0^2)^2+2\,(f_0+1)^2\right]+
\frac{3}{16\,t^{2\,f_0+4}x^{2\,k}}
\left[(1-p_0)^2(1+f+p)^2\right.\nonumber\\
&&\left.+(1+p)^2(1+f-p)^2 \right]
\,.\ea
Since (\ref{condi}) holds it can be seen that $K$  is singular at
$t=0$ for any value of the 
three free parameters of the solution, indicating thus
the generic presence of a spacelike singularity at the time origin.
 
It is also possible to study the singularity structure of the solutions
in  a more refined  way, in particular by studying 
the expansion along each spatial axis. In general, the behavior
will strongly depend on the values of the parameters $k$, $C$ and
$\beta$. It can be shown analytically that for large enough $\beta$ 
there will necessarily be contraction along the $z$ axis, moreover if
$k>1$ that will be the case regardless the value of $\beta$ and $C$. 
Another fact that can be easily checked is the impossibility 
of having simultaneous contraction along axes $x$ and $y$.
Three main types of singular behavior can be distinguished:

\indent{\bf a) Point-like singularities (Quasi-Friedmann behavior)}\\
All three spatial directions shrink as the initial
time $t=0$ is approached; 
or explicitly $lim_{t\rightarrow 0} \,\epsilon_i=0\; \forall i$.
Depending on how many directions have the same
expansion rate the behavior will be completely anisotropic, 
axially symmetric or isotropic.

\indent{\bf b) Finite lines}\\
This type of singular behavior occurs when in the vicinity of $t=0$ one 
of the spatial directions neither expands nor contracts with time; 
in other words, it is said that the direction $i$ is a finite line if 
$lim_{t\rightarrow 0} \,\epsilon_i=1$. The subcases can be classified 
according to which the direction behaves in that way. 
In general, there will be a single finite line, though
it is possible to have particular cases in which a second finite 
line exists. However, it is impossible to have such behavior along all
three directions.

\indent{\bf c) Infinite lines (Quasi Kasner regime)}\\
An infinite line along the $i$ direction exists when 
$lim_{t\rightarrow 0} \,\epsilon_i=\infty$. Again, three cases can be seen
to occur, depending on which is the axis displaying that feature.
For some particular values of the parameters
the maximum allowed number of two infinite lines
can be reached.

Since the singular behavior of the metrics depends on three
parameters it becomes rather complicated to represent 
graphically the structure of singularities in the general case. 
For that reason I will restrict myself to cases in which the
value of one of the parameters is fixed, namely $C$.
In the pictures here two different sets of lines can be distinguished.
On the one hand, the black  lines in each plot correspond to the
curves along which a  spatial direction becomes a finite line.
In the region delimited by the black continuous line and the axes, 
a infinite line type of singularity arises along direction $z$.
Here one can  see how for $C=0.5$  it is possible to have simultaneously 
the same behavior along directions $y$ and $z$.
For $k>1$ in the region above the black dashed line there 
is contraction along direction $y$, this behavior gets reversed for $k<1$.
Similarly, for $\beta$ values less than those along the dashed-dotted
line contraction takes place, and the contrary happens for $k>1$.
The points where two black lines intersect correspond to having
two finite lines. On the other hand, the grey lines represent the curves
along which two spatial directions display  the same
expansion rate. The fact that at a given point
the three grey lines intersect
reflects the possibility of having isotropic expansion, 
and for the two $C$ values chosen, that point lies in the $k>1$ region.  
\begin{figure}[ht!]
\centering
\leavevmode
\epsfxsize=3in
\epsfbox{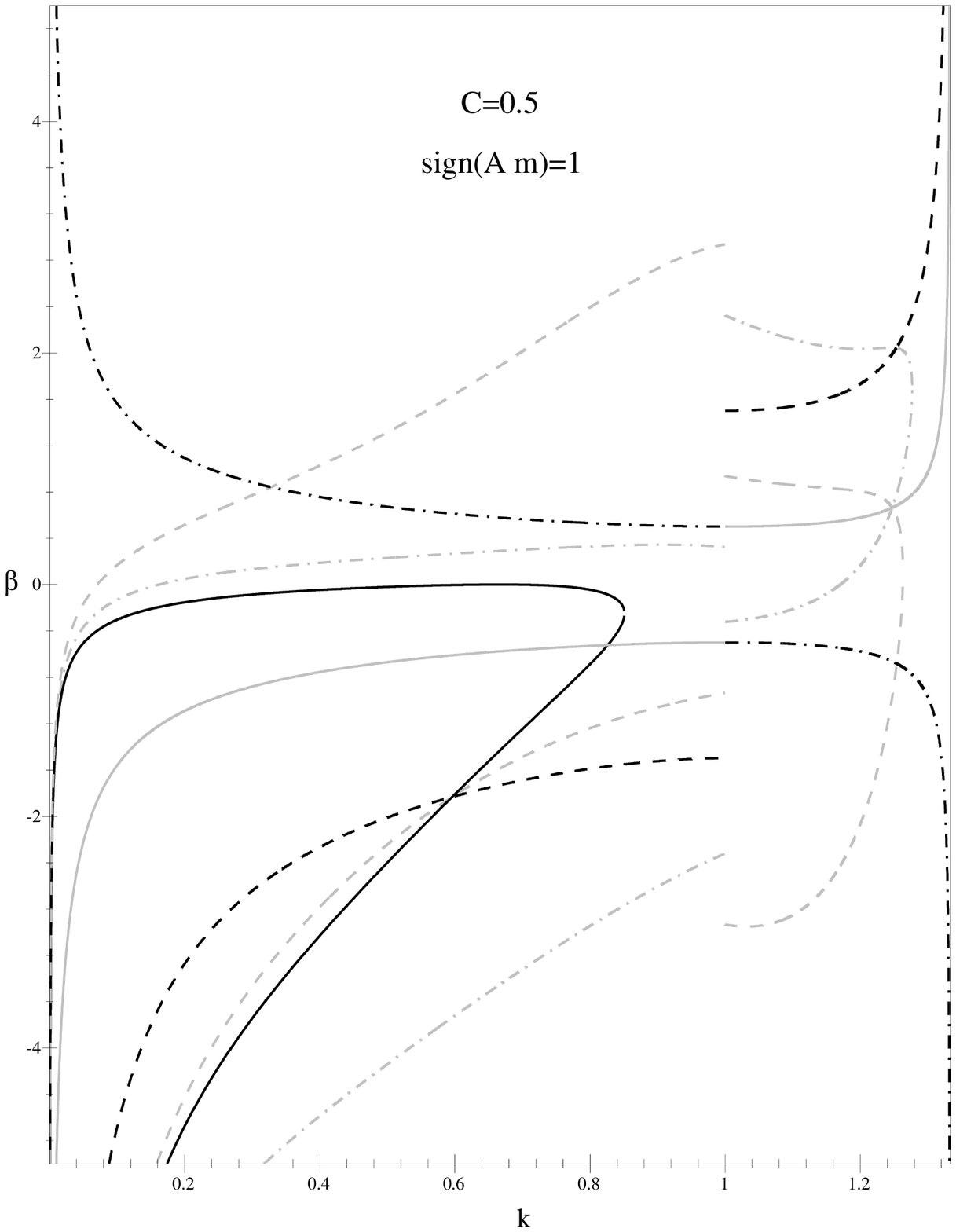}
\leavevmode
\epsfxsize=3in
\epsfbox{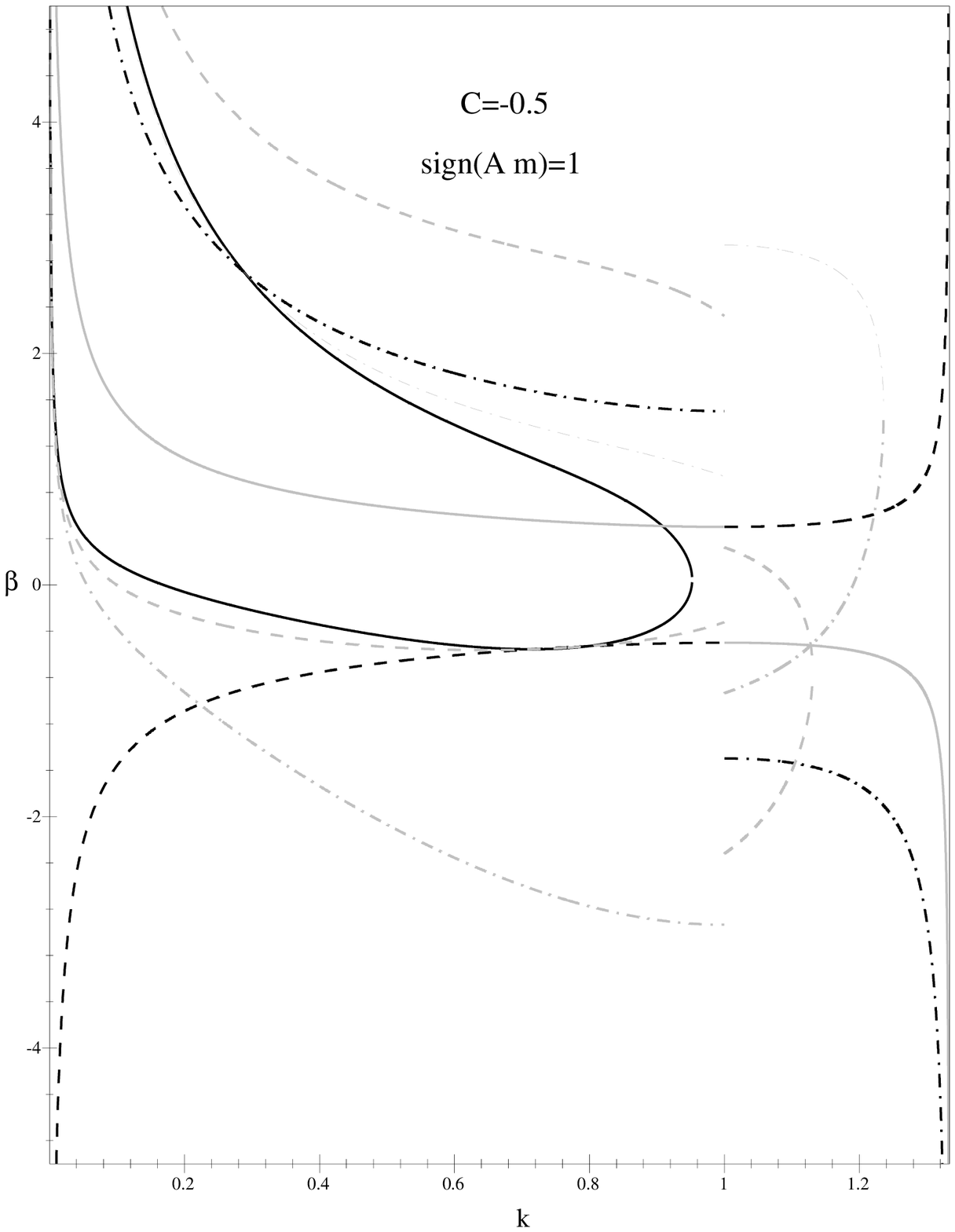}
\caption{Structure of singularities of the 
inhomogeneous generalization of Belinskii-Khalatnikov's model for
$C=0.5$ (l.h.s. figure) and $C=-0.5$ (r.h.s. figure) with 
$\hbox{sign}{\,(A\,m)}$.
On the one hand, the black lines indicate the value of $\beta$ as a function
of $k$ for which a given direction behaves as a finite line.
The black continuous, dashed and dashed-dotted lines correspond to 
a finite line along $z$, $y$ or $x$ directions respectively.
On the other hand, the grey lines indicate the value 
of $\beta$ as a function
of $k$ for which two spatial directions have the same 
expansion rate. The grey continuous, dashed and 
dashed-dotted lines correspond respectively to 
$\epsilon_2=\epsilon_3$, $\epsilon_1=\epsilon_3$ 
and  $\epsilon_1=\epsilon_2$.}
\end{figure}

\subsection{WKB limit}
With regard to the physical interpretation of the new 
solutions constructed here,
it is their high-frequency limit which turns out to be most
interesting.  This limit, also called the WKB regime,
 corresponds to the regime in  which the time elapsed since the beginning
of the Universe is much larger than the period of any perturbation 
mode. By taking $j\,t\gg 1$, for every value of $j$, in the normal mode 
expansions of the scalar field and metric functions, 
Charach and Malin were able to show that the
relativistic solutions taken here as input, represent scalar and
gravitational waves propagating on an spatially flat background. In this
limit, such universes are causally connected
because the particle horizon is larger than the wavelength of any of the
modes of the independent degrees of freedom, namely the transverse
part of the gravitational field $p$ and the scalar field $\tilde \varphi$.

In this vein, it will be proved here in two different ways 
that the $G_1$ counterparts to  Charach and Malin's 
cosmological models can also be thought of in terms of waves
propagating on a non-static background. Thus,  the physical interpretation
is not spoiled in the process of homogeneity breaking. A direct 
consequence of the additional degree of inhomogeneity present in the $G_1$ 
solutions, as compared to their more symmetric counterparts, is that 
the background on which the waves live  is not spatially flat. 
Cosmological backgrounds perturbed by waves  have also been considered 
in a series of paper by Centrella and Matzner \cite{matz1,matz2,cen}
who studied collisions of plane gravitational waves in these settings.

Let us consider now the late time expressions for the metric functions and the 
scalar field of the $G_1$ spacetimes obtained by applying the new
technique to the cosmological models given by expressions (\ref{full}):
\bml\ba
p&\sim& p_0\log t+B\,\bar p(t,z)\,,\\
f&\sim& f_0\log t +
\left(\frac{t}{2\pi}\right) \sum_{j=1}^{\infty}j \alpha_j^2\,,\\
\tilde \varphi&\sim&\varphi_0\log t+
A\,\bar p+m \log \vert x \vert\,,\\
\bar p&=&\sum_{j=1}^{\infty}\sqrt{\frac{2}{j\pi t}}\alpha_j 
\cos (j\, t-\frac{\pi}{4}) \cos [j (z-z_j)]\,.
\ea\eml
Since in this regime $\bar p\ll 1$, the metric $\tilde g_{\mu\nu}$ 
can be split into a background $\tilde \eta_{\mu\nu}$ 
plus a perturbation metric $\tilde h_{\mu\nu}$, that is:
\bml\ba
\tilde g_{\mu\nu}&\sim& \tilde \eta_{\mu\nu}+\tilde h_{\mu\nu}\,,\\
\tilde \eta_{\mu\nu}&=&\hbox{diag}
\left(-x^kt^{f_0}e^{f_1\,t},x^kt^{f_0}e^{f_1\,t},
t^{1+p_0},x^nt^{1-p_0}\right)\,,\\
\tilde h_{\mu\nu}&=&\hbox{diag}
\left(0,0, t^{1+p_0}B\,\bar p,-x^nt^{1-p_0}B\,\bar p\right)\,.
\ea
\eml
Here, in addition to (\ref{cons}), I have made the following definition
\footnote{Though the factor $B^2+2\,A^2$ equates to unity in the simple
case I am dealing with, it has been deliberately introduced 
in the definition of $f_1$; so that the trail of the separate  
contributions to the energy momentum tensor
of the graviton and scalar field pair can be followed. 
Had I considered the general case of the procedure to generate 
a $G_2$ massless scalar field solution, then 
$ f_1\ne\sum_{j=1}^{\infty}{j \alpha_j^2}{/(2\,\pi)}\,.$}:
\be  f_1=\sum_{j=1}^{\infty}\frac{j \alpha_j^2}{2\,\pi}\,(B^2+2\,A^2)\,.\ee

A peculiarity regarding the perturbations on the scalar and
gravitational degrees of freedom is that for $0<k<1$  they  
will be on phase whereas for $1<k<4/3$ they  will be phase-shifted by $\pi$.
Nonetheless, it will be seen later that whatever the value of $k$ the
scalar and gravitational perturbations contribute constructively to 
the energy momentum tensor.  
  
A proof of the assertion made above regarding the non spatial flatness of 
the spacetimes described by the background metric $\tilde
\eta_{\mu\nu}$, 
is provided by the expression of the spatial curvature of the 
$t=constant$ three dimensional hypersurfaces:
\be^{(3,\eta)}R=\frac{m^2}{2\,x^2t^{p_0+2 }}.\ee

Let us now proceed to analyze the $\tilde h_{\mu\nu}$ 
tensor and thereby show that it represents the gauge-invariant perturbations 
of a background spacetime $\tilde \eta_{\mu\nu}$;
or in other words, that it satisfies the wave equation
$\tilde\nabla^{\gamma}\tilde\nabla_{\gamma} \tilde h_{\mu\nu}=0$, 
or equivalently, following \cite{muk}, that:
\ba &\tilde h _{\,\gamma}^{\gamma}=0& \label{traza}\,,\\
& \tilde\nabla ^\gamma \tilde h_{\gamma\lambda}=0& \label{trans}\,,\ea
with $\gamma,\lambda=2,3$, and where the D'Alambertian and the 
covariant derivative must be calculated using the corresponding 
background metric $\tilde \eta_{\mu\nu}$. 

Since it is straightforward to see that the trace-free condition
(\ref{traza}) is satisfied by construction, the problem reduces
to satisfying (\ref{trans}), which is equivalent 
to requiring:
\begin{mathletters}
\label{wave}
\ba
&\,&\tilde h_{22,x}-2\,\tilde\Gamma_{22}^2 \tilde h_{22}=0 \label{waveuno}\,,\\
&\,&\tilde h_{33,x}-2\,\tilde\Gamma_{32}^3 \tilde h_{33}=0 \label{wavedos}\,.
\ea
\end{mathletters}
On the one hand, expression (\ref{waveuno}) 
is identically null because neither $\tilde \eta_{22}$ nor 
$\tilde h_{22}$ are $x\hbox{-dependent}$.
On the other hand, it can be seen that (\ref{wavedos}) is also satisfied
by just having in mind that: 
\bml
\ba
&&\tilde h_{33}=B\,\tilde\eta_{33}\,,\\
&&\tilde\Gamma_{32}^3=\left(\log\sqrt{\tilde \eta_{33}}\right)_{,x}\,.
\ea
\eml
Once it  has been proved that the wave equation
 $\tilde\nabla^{\gamma}\tilde\nabla_{\gamma} \tilde h_{\mu\nu}=0$ holds 
one  can properly refer to the tensor $\tilde h_{\mu\nu}$ 
as describing metric perturbations in the form of gravitational waves.

In order to give additional arguments in favor of the interpretation of the
solution in terms of waves propagating on a non flat background, 
I shall follow Charach and Malin \cite{cha} and 
analyze the energy-momentum tensor of
the background metric $\tilde\eta_{\mu\nu}$. It will be shown
that the stress-energy tensor is naturally manifested in
terms of two components.
One of these corresponds to a null fluid, supporting thus
the interpretation suggested above; while the other term 
corresponds to an inhomogeneous
massless scalar field with no $z\hbox{-dependence}$. 
In particular, 
\be
^{(\eta)}\tilde T_{\mu}^{\nu}=^{(1)}\tilde T_{\mu}^{\nu}+
^{(2)}\tilde T_{\mu}^{\nu}\,,
\ee
where
$^{(1)}\tilde T_{\nu}^{\nu}\ne 0$ and $^{(2)}\tilde T_{\nu}^{\nu}=0$.
Explicitly
\bml\ba ^{(1)}\tilde T_0^0&=&-{t^{-(1+{p_0})}}\frac{m^2 \,
      }{2\,{x^2}}} - {t^{-(2+{f_0})}}{\frac{1 + 2\,{f_0} - 
        {{{p_0}}^2}}{4\,{e^{{f_1}\,t}}\,{x^k}}\,,\\
^{(1)}\tilde T_1^1&=& -{t^{-(1 + {p_0})}}\frac{m^2 \,
      }{2\,{x^2}}} + 
  t^{-(2 + {f_0})}{\frac{ 1 + 2\,{f_0} - 
        {{{p_0}}^2}\,
      }{4\,{e^{{f_1}\,t}}\,{x^k}}\,,\\
^{(1)}\tilde T_2^0&=&{t^{-(1 +{f_0})}} \frac{ 2\,{p_0}\,\left( k-1\right) 
         -k \,}{2\,
     {e^{{f_1}\,t}}\,{x^{1 + k}}}\,,\\
^{(1)}\tilde T_2^2&=&  {t^{-(1+{p_0})}}\frac{m^2 \,
        }{2\,{x^2}}} + 
  {t^{-(2 + {f_0})}} {\frac{ 1 + 2\,{f_0} - 
        {{{p_0}}^2} }{4\,
      {e^{{f_1}\,t}}\,{x^k}},\\
^{(1)}\tilde T_3^3&=&-  {t^{-(1 + {p_0})}}\frac{m^2 \,
     }{2\,{x^2}}} + 
  {t^{-(2 + {f_0})}} {\frac{ 1 + 2\,{f_0} - 
        {{{p_0}}^2}  \,}{4\,
      {e^{{f_1}\,t}}\,{x^k}}\,,\ea
\ba
 ^{(2)}\tilde T_0^0&=&-t^{-(1+f_0)}\frac{f_1}{2\,e^{f_1t}\,x^k}\,,\\
^{(2)}\tilde T_1^1&=&t^{-(1+f_0)}\frac{f_1}{2\,e^{f_1t}\,x^k}\,. 
\ea\eml 

I will now proceed to give an interpretation for the
$^{(1)}\tilde T_{\mu}^{\nu}$ term. The Klein-Gordon equation for a scalar
 field $\tilde \psi$, calculated using the metric 
$\tilde\eta_{\mu\nu}$, takes the form
\be 
\frac{(x\tilde \psi_{,x})_{,x}}{x^{1-k}}
-\frac{(t\,\tilde\psi_{,t})_{,t}}{e^{f_1 t}t^{f_0-p_0}}=0,\ee
a solution of which is  
\be \tilde\psi=\varphi_0\log t+m\log \vert x \vert. \ee
The energy momentum tensor for the field $\tilde \psi$
propagating on the spacetime $\tilde \eta_{\mu\nu}$ 
yields $^{(1)}\tilde T_{\mu}^{\nu}$ exactly. That being so, one can conclude
that the exact solution obtained after applying the 
generating technique to
(\ref{full}) asymptotically evolves into a solution with a single
degree of inhomogeneity; that is, the sinusoidal inhomogeneities along 
the $z\hbox{-axis}$ vanish with time. In other words, 
$^{(1)}\tilde T_{\mu}^{\nu}$  corresponds to the energy-momentum
tensor of the exact solution  one would obtain by 
switching off the periodic inhomogeneities along $z$. 
 
On the other hand, the traceless  term $^{(2)}\tilde
T_{\mu}^{\nu}$ can be shown to account for waves. 
It can also be separated into two parts, namely:
\bml\ba
^{(2)}\tilde T_{\mu\,\nu}&=&^{(GW)}\tilde T_{\mu\,\nu}+
^{(SW)}\tilde T_{\mu\,\nu},\\
^{(GW)}\tilde T_{\mu\,\nu}&=&\sum_{j=-\infty}^{\infty}
\frac{B^2\,\alpha_j^ 2}{4\,|j|}{\kappa}_{\mu j}{\kappa}_{\nu j},\label{gw}\\
^{(SW)}\tilde T_{\mu\,\nu}&=&\sum_{j=-\infty}^{\infty}
\frac{A^2\,\alpha_j^ 2}{2\,|j|}{\kappa}_{\mu j}{\kappa}_{\nu j}\label{sw}
.\ea\eml
The null vector $\kappa_{\mu}$ is defined by
\be
\kappa_{\mu j}=\frac{1}{\sqrt{\pi\,t}}\,(\vert j\vert,j,0,0).\ee

It is clear that $^{(2)}\tilde T_{\mu\nu}$
corresponds to  a null fluid describing a collisionless
flow  of ``gravitons" and ``scalar particles". 
It is easy to see that in the case under discussion 
gravitational waves will be absent only if the metric has no 
$x\hbox{-dependence}$.

I has been shown that the new solutions represent, in their
WKB limit, waves propagating on an inhomogeneous non-static 
spacetime. In our case the only difference with respect 
to the case studied by Charach and Malin  is that the 
background is not spatially flat. 

As far as the  further evolution of the model is concerned, 
the presence of the additional degree of inhomogeneity in 
the model plays a crucial role. The null fluid's contribution 
to the energy-momentum tensor dominates
the one due to the homogeneous part of the scalar field, and
in the $t\gg 1$ limit the energy-momentum tensor
has a term accounting for the waves, plus another  
corresponding to the $x\hbox{-dependent}$ term in the scalar field.
The scalar curvature of the background in this limit will be given by:
\be ^{(\eta)}R\sim\frac{m^2}{x^2\,t^{1+p_0}},\ee
so that this universe can either become singular at $t=\infty$  or flat,
depending on the sign of $1+p_0$. The possibility of having a 
curvature spacelike singularity at late times is 
entirely due to the inhomogeneous character of the background. 

Another proof of the interpretation of the solutions in terms of a 
null fluid propagating on the background $\tilde \eta_{\mu\nu}$
can be given. Let us calculate the energy- momentum tensor that
corresponds to the scalar field $\tilde\varphi$ in the late time 
limit and just retain terms up to the order $t^{-1}$. Under
this restriction the only non-null terms of the energy momentum tensor 
are $\tilde T_{00}$ and $\tilde T_{11}$, which are given by:
\ba
&&\tilde T_{00}=-\tilde T_{11}=\left(\frac{A^2}{\pi\,t}\right)
\sum_{l=1}^{\infty}\sum_{j=1}^{\infty}
{\alpha_l \alpha_j\sqrt{lj}} \left\{\sin [l(z-z_l)] 
\sin[j(z-z_j)]\cos[l\,t-\frac{\pi}{4}]\cos[j\,t-\frac{\pi}{4}]\right.\nn
&&
-\cos [l(z-z_l)]\cos[j(z-z_j)]
\left. \sin[l\,t-\frac{\pi}{4}]\sin[j\,t-\frac{\pi}{4}]\right\}+
{\cal O}(t^{-2}).\ea
Averaging $\tilde T_{00}$ over a region $0 \leq t\leq 2\,\pi$, 
$0 \leq z\leq 2\,\pi$ on the $(t,z)\hbox{-plane}$ one obtains:
\be\langle \tilde T_{00}\rangle=-\langle \tilde T_{11}\rangle=
\frac{1}{4\,\pi^2}\int_{0\,}^{2\,\pi}\int_{0\,}^{2\,\pi}
\langle \tilde  T_{00}\rangle \,dzdt=\frac{1}{2\,\pi\,t}
\sum_{j=1}^{\infty}j\,A^2.\ee
In the same approximation it can be seen that
\be \langle \tilde T \rangle=\langle \tilde T_2^2 \rangle=
\langle \tilde T_3^3\rangle ={\cal O}(t^{-2}).\ee
So, essentially it has been found that
\be\langle \tilde T_{\mu\nu} \rangle=^{(SW)}\tilde T_{\mu\nu}.\ee
Summarizing, both methods have yielded the result that for $t \gg 1$, 
and up to the order $t^{-1}$, 
the energy-momentum of the m.s.f. can be reduced to the null fluid form.  

The WKB regime of the solutions under discussion admits a 
reformulation in terms of the density of particles contributing 
to the modes of two fields. I shall strictly follow Charach's 
approach here \cite{car}, which consists of performing a quasi-classical
treatment based on the geometrical optics energy-momentum tensor.
A family of Lorentz local frames is introduced so that the 
density of particles in each normal mode can be defined through:
\be\lambda^{(a)} =\sqrt{\vert\bar \eta_{\nu \nu} \vert}\,\delta_{\nu}^{(a)} \;\;, 
\hbox{(no summation over } \nu \hbox{)}
\ee
where $\bar \eta_{\nu \nu}$ represents the components of 
the inhomogeneous generalization 
of the DZN metric. A set of observers corresponding
to this tetrad are characterized by the 4-velocity:
\be
u^{\nu}\equiv \lambda_{(0)}^{\nu}=\left(\sqrt{\bar \eta^{00}},0,0,0 \right) .
\ee
Let us consider now equations (\ref{gw},\ref{sw}),
which give the WKB stress-energy tensors of 
the cosmological model with gravitational and scalar wave
perturbations, namely:
\bml\ba
^{(GW)}\tilde T_{\mu\,\nu\,j}&=&
\frac{B^2\,\alpha_j^ 2}{4\,|j|}{\kappa}_{\mu\,j}{\kappa}_{\nu\,j}\\
^{(SW)}\tilde T_{\mu\,\nu\,j}&=&
\frac{A^2\,\alpha_j^ 2}{2\,|j|}{\kappa}_{\mu\,j}{\kappa}_{\nu\,j}
.\ea\eml
The density of scalar and gravitational particles in the $n\hbox{-th}$ 
mode is given by
\bml\ba
\rho_j^S&=&\frac{\tilde T_{(0)j}^{(0)SW}}{h\kappa_{\mu j}^{(0)}}=
\frac{A^2\,\alpha_j^ 2}{2\,t\,\sqrt{x^{k}\,e^{f}}}\\
\rho_j^G&=&\frac{\tilde T_{(0)j}^{(0)GW}}{h \kappa_{\mu j}^{(0)}}=
\frac{B^2\,\alpha_j^ 2}{4\,t\,\sqrt{x^{k}\,e^{f}}},
\ea\eml
where $\kappa_{\mu \nu}$ is a null vector with dimensions of length. 
Besides, since the description here is based on units $c=G=1$, 
the Planck constant has dimensions 
of $\hbox{(length)}^2$, where $h\sim 10^{-66} cm^2$.
The density does not depend on the direction along which 
the particles propagate, that is:
\bml\ba \rho_j^S&=&\rho_{-j}^S\\
\rho_j^G&=&\rho_{-j}^G.\ea\eml
In the $G_1$ models considered here, the volume of the 
spatial sections $t=constant$  is not finite; for that reason the
total number of particles in each mode
of the two degrees of freedom has no upper bound.

In light of this reformulation it was suggested that one should
regard the evolution of $G_2$ models filled with waves  
as describing a process of transforming the initial 
inhomogeneities along $z$ into quanta of various fields. Clearly,
this interpretation's  validity is extendible to models
with just one  isometry, such as those constructed in this paper.
\section{Conclusions}
Before I finish, I will summarize the main results. I have presented the
first method to generate uniparametric families of general relativistic
cosmologies having a two-dimensional inhomogeneity and a m.s.f. as a source. 
In the context of either General Relativity or alternative theories
of gravity, one can generate a large number of new inhomogeneous 
cosmologies using this algorithm, where moreover the
only input needed is any of the many known
vacuum relativistic cosmologies with two 
commuting Killing vectors.

It has also been shown that this technique allows one to
construct families of cosmologies which represent waves propagating on
a spatially curved cosmological background. 
The spacelike singularity structure of these solutions 
has been studied, and several peculiarities due
to the matter content have been elucidated.  
\section*{Acknowledgements}
I am much indebted to Alexander Feinstein, 
Miguel \,Angel V\'azquez-Mozo and Dominic Clancy for valuable suggestions and
enlightening discussions. Useful comments 
from Malcolm MacCallum, Bruce Bassett, 
Gary Kerr and Juan Antonio Valiente are  also acknowledged. 
This work has been carried out thanks to the financial support of the Basque
Government under fellowship number BFI98.79 and of the Spanish Ministry
of Education and Culture  under grant 172.310-0250/96.



\end{document}